\documentstyle[aps,manuscript]{revtex}

\begin{document}
\title{Lagrangian approach to a symplectic formalism \\
for singular systems}
\author{H. Montani and R. Montemayor}
\address{Centro At\'{o}mico Bariloche, CNEA \\
and \\
Instituto Balseiro, Universidad Nacional de Cuyo \\
8400 - S. C. de Bariloche, R\'{\i}o Negro, Argentina}
\maketitle

\begin{abstract}
We develop a Lagrangian approach for constructing a symplectic structure for
singular systems. It gives a simple and unified framework for understanding
the origin of the pathologies that appear in the Dirac-Bergmann formalism,
and offers a more general approach for a symplectic formalism, even when
there is no Hamiltonian in a canonical sense. We can thus overcome the usual
limitations of the canonical quantization, and perform an algebraically
consistent quantization for a more general set of Lagrangian systems.
\end{abstract}
\vskip 0.5cm

\section{Introduction}

There is a canonical recipe for accomplishing the conventional quantization
of a dynamical system. From a given Lagrangian we use a Legendre
transformation to rewrite the velocities in terms of the momenta and to
determine the Hamiltonian, and next from the Poisson brackets between
coordinates and momenta we define the quantum commutation relations. This
procedure works reasonably well for regular Lagrangians, but it becomes
troublesome for the singular ones, which are characterized by the
non-invertibility of their Hessian matrix. This not only prevents us from
determining all the accelerations in terms of the coordinates and their
velocities, but it also confronts us with a singular Legendre transformation
when we try to construct the canonical theory. These difficulties reflect the
use of an excess of coordinates in the original description, because the true
physical degrees of freedom are contained in a subspace of the configuration
space.  This subspace is defined by a set of Lagrangian constraints, and its
complement is a set of physically equivalent configurations or a physically
inaccessible region.

Some decades ago, Dirac and Bergmann and colaborators made a very relevant
contribution to the subject of singular Lagrangians, and developed a
systematic way of constructing a canonical formalism for these systems\cite
{DB,D,SM,S}. The Dirac method has played a crucial role in the quantization
of many theories of significant physical relevance, as is the case of gauge
theories and their gauge fixing procedure. However, there exists some
pathological examples where the Dirac method seems to fail\cite{C,F},
apparently because the constraint analysis on which it is based is not
exhaustive enough\cite{GNH,LL}. In these cases the Dirac construction of the
phase space leads to some inconsistencies in relation to the original
Euler-Lagrange equations. The meaning of the primary/secondary
classification of the constraints is not clear\cite{SM,S}, and the
connection between the constraints and the symmetries of the system is
rather vague. This has been manifested by a large body of work speculating
on the use of full or partially extended Dirac Hamiltonians, following more
or less ad hoc criteria\cite{C,F,GNH}. Besides this, for Lagrangians which
are not bilinear functions of the velocities, a one to one relation between
both velocities and momenta might not exist. This introduces some additional
problems with the use of the Legendre transformation and spoils the
construction of a canonical classical formalism and of a consistent
canonical quantization \cite{MH,HT}. If we consider the Lagrangian
description for a dynamical system as the basic one, these problems are
related to the fact that the Dirac approach gives a set of constraints which
are not unambiguously defined and with no direct relation with the
Lagrangian constraints.

The Dirac approach is an extension of the most usual method of constructing
a canonical formalism for regular systems, based on a Legendre
transformation. But this is not the only approach to achieve this aim. One
possiblity, based on the theory of transformations for the Lagrangian
formalism, has been explored in Ref. \cite{BG}. Another possibility is to
visualize the canonical formalism for regular systems as a very special
first order Lagrangian theory. This first order theory is constructed from
the original one by redefining the velocities as new auxiliary variables,
which leads to a new configuration space of double dimension.This is the
perspective that we exploit in the present work, which allows us to obtain a
canonical formalism maintaining a clear relationship with the original
Lagrangian, and avoiding the ambiguities involved in the usual construction
of the phase space.

Here we develop, in a heurisitic language and using local coordinates, such
an alternative procedure. It is based on a redefinition of the original
velocities as new variables, leading to a Lagrangian with a linear
dependence on the new velocities in a configuration space of double
dimension. Essentially, it is an extended version for singular systems of
the approach discussed by Lanczos\cite{L} to accomplish the transition from
the Lagrangian to the Hamiltonian form, which is based solely on the method
of Lagrangian multipliers. In this context we construct a symplectic
structure for singular Lagrangians. The use of Lagrange multipliers allows
us to overcome the sometimes not well defined Legendre transformation.

A few years ago, Faddeev and Jackiw\cite{FJ} proposed a method, based on the
classical geometrical approach to dynamical systems, in order to identify
cyclic variables and the associated constraint functions. They then
proceeded to use them to eliminate superfluous degrees of freedom. However,
it sometimes happens that such an elimination is not convenient or even
possible. To handle these situations, these constraints have been
incorporated into the kinetic terms by using velocities as Lagrange
multipliers\cite{BC}. This procedure has been shown to be equivalent to the
Dirac method when the latter is applicable\cite{M}, and offers a clear
description of the gauge symmetries \cite{CM}. Here we apply the schema
presented in Ref. \cite{BC,M,CM} to develop a transparent analysis of the
structure of singular Lagrangians and their constraints in a generalized
symplectic context. In particular, this analysis sheds light on the Dirac
conjecture, i.e., on the relationship of first class secondary constraints
with gauge symmetries\cite{CM}.

A very important feature of our approach is that the resulting symplectic
formalism guarantees the soundness of the original Lagrangian equations of
motion. We will deal here with finite-dimensional systems, but the main
results can be extended without difficulty to field theory.

The development of the article is as follows. In Section II we discuss how
to construct from the original Lagrangian a first order one that leads to a
presymplectic structure. If the original Lagrangian is regular, at this
stage we already have the Poisson brackets and the symplectic formalism. In
the three following sections we follow the analysis for singular Lagrangians
and discuss the appearance of constraints. Here we use the derivative
Lagrange multipliers to incorporate the constraints to the first order
Lagrangian and study the different cases. The resulting schema allows us to
give a complete characterization of the symmetries in relation to the
constraints, and to construct a well defined symplectic formalism. To
illustrate our approach, Section VI contains two examples, already discussed
in the literature as presenting a pathological character from the Dirac
point of view. Here we show how our approach allows us to treat them.
Finally, the last section is devoted to some comments and conclusions.

\section{The primary first order Lagrangian and its constraints}

Let us consider a dynamical system described by a Lagrangian $L(q_i,\dot{q}
_i)$, of arbitrary order in the time derivatives $\dot{q}_i$, $i=1,...,n$.
To develop a symplectic form for this system, we start by transforming this
Lagrangian to an adequate first order one, by extending the schema of Lanczos
\cite{L} to singular Lagrangians. The procedure consists in substituting in
the Lagrangian the velocities $\dot{q}_i$ by $n$ auxiliary variables $\alpha
_i$ to define
\begin{equation}
\hat{L}(q_i,\alpha _i)=L(q_i,\dot{q}_i)|_{\dot{q}_j=\alpha _j}\;,
\end{equation}
and in constructing a new Lagrangian $L_{_{(0)}}(q,\dot{q},\pi ,\alpha )$,
by adding to $\hat{L}(q,\alpha )$ the constraints $\alpha _i=\dot{q}_i$
through a set of Lagrange multipliers $\pi _i$:
\begin{equation}
L_{_{(0)}}(q,\dot{q},\pi ,\alpha )=\pi _i(\dot{q}_i-\alpha _i)+\hat{L}
(q,\alpha )\;.
\end{equation}
This new Lagrangian leads to exactly the same equations of motion for the
$q_i$ variables than the original $L(q_i,\dot{q}_i)$, as can be easily
verified by eliminating the cyclic variables $(\pi _i,\alpha _i)$\cite{EE}.
The structure of $L_{_{(0)}}$ becomes more meaningful when isolating the
velocity-independent terms:
\begin{equation}
L_{_{(0)}}(q,\dot{q},\pi ,\alpha )=\pi _i\dot{q}_i-H(q,\pi ,\alpha )\;.
\label{L0}
\end{equation}
The function of the extended configuration space
\begin{equation}
H(q,\pi ,\alpha )=\alpha _i\pi _i-\hat{L}(q,\alpha )  \label{H}
\end{equation}
is a constant of motion for the Lagrangian $L_{_{(0)}}(q,\dot{q},\pi ,\alpha
)$. The $\alpha _i$ are cyclic variables, and their equations of motion give
rise to a set of primary constraints:
\begin{equation}
\Omega _{_{(1)}}^i\equiv -\frac{\partial H(q,\pi ,\alpha )}{\partial \alpha
_i}=\frac{\partial \hat{L}(q,\alpha )}{\partial \alpha _i}-\pi _i=0\;.
\label{PC}
\end{equation}
These constraints resemble the definition of the canonical conjugate
variables in the usual Legendre transformation, but here they are defined in
an extended configuration space, and simply characterize its enlargement.
Expression (\ref{L0}) also strongly resembles the usual relation between
the Lagrangian and the Hamiltonian function but here, due to the
constraints, the equations of motion cannot be cast into a canonical form
and it is not possible to define a kind of generalized Poisson brackets.

On the subspace defined by (\ref{PC}) the function $H$ is $\alpha $
-independent. If the constraint $\Omega _{_{(1)}}^i=0$ allows us to
algebraically explicitate $\alpha _i$, it can be used to eliminate this
variable from the Lagrangian $L_{_{(0)}}$\cite{S}, and thus to reduce the
dimension of the extended configuration space. By doing this we show in
Appendices A and B how the canonical and the Dirac Hamiltonian appear.

In the following, to make clear the role of the Lagrangian constraints, we
will maintain the whole set of auxiliary variables. To promote the
constraints to Euler-Lagrange equations they must be incorporated to the
Lagrangian $L_{_{(0)}}(q,\dot{q},\pi ,\alpha )$ by means of Lagrange
multipliers. We characterize the constraints by their time derivatives,
instead of the $\Omega _{_{(1)}}^i$ constraints themselves\cite{BC,M}. This
is easily implemented by introducing velocities as Lagrange multipliers,
$\dot{\lambda}_{_{(1)}}^i$, in place of the standard ones. This procedure has
the advantage that only the velocity-dependent term is modified, while the
velocity-independent part remains the same. Thus, from here on the resulting
function $H(q,\pi ,\alpha )$ is completely defined, and the insertion of the
new constraints will only modify the coefficients of the velocities in the
equations of motion. In such a way we obtain a new Lagrangian $L_{_{(1)}}(q,
\dot{q},\pi ,\alpha ,\dot{\lambda}_{_{(1)}})$:
\begin{equation}
L_{_{(1)}}(q,\dot{q},\pi ,\alpha ,\dot{\lambda}_{_{(1)}})=\pi _i\dot{q}
_i+\Omega _{_{(1)}}^i\dot{\lambda}_{_{(1)}}^i-H(q,\pi ,\alpha )\;.
\end{equation}

For the sake of brevity, let us introduce the compact notation
$Q_{_{(1)}}^{^A}=(q;\pi ;\alpha ;\lambda _{_{(1)}})$, $\Delta _{_{(1)}}^{^A}
={\frac{\partial L_{_{(1)}}}{\partial \dot{Q}_{_{(1)}}^{^A}}}$, where the
range of the $A$ index is $\{1...n;1...n;1...n;1...n\}$, such that
$L_{_{(1)}}$ may be written as
\begin{equation}
L_{_{(1)}}(Q_{_{(1)}})=\Delta _{_{(1)}}^{^A}\dot{Q}
_{_{(1)}}^{^A}-H(Q_{_{(1)}})\;,
\end{equation}
with $H(Q_{_{(1)}})=H(q,\pi ,\alpha )$ given by Eq. (\ref{H}). This
Lagrangian leads to the first order equations of motion:
\begin{equation}
F_{_{(1)}}^{^{AB}}\dot{Q}_{_{(1)}}^B=-\frac{\partial H(Q_{_{(1)}})}{\partial
Q_{_{(1)}}^A}\;,
\end{equation}
where $F_{_{(1)}}^{^{AB}}={\frac{\partial \Delta _{_{(1)}}^{^B}}{\partial
Q_{_{(1)}}^{^A}}}-{\frac{\partial \Delta _{_{(1)}}^{^A}}{\partial
Q_{_{(1)}}^{^B}}}$ has the explicit form
\begin{equation}
F_{_{(1)}}=\left(
\begin{array}{ccc}
f & 0 & M_{_{(1)}} \\
0 & 0 & R_{_{(1)}} \\
-M_{_{(1)}}^{^T} & -R_{_{(1)}}^{^T} & 0
\end{array}
\right) \;.
\end{equation}
The matrix $f$ is the symplectic matrix in the $(q,\pi )$ space
\[
f=\left(
\begin{array}{cc}
0 & -I \\
I & 0
\end{array}
\right) \;,
\]
where $I$ is the unit matrix, and the remaining components are
\begin{equation}
M_{_{(1)}}=\left(
\begin{array}{c}
{\frac{\partial \Omega _{_{(1)}}}{\partial q}} \\
{\frac{\partial \Omega _{_{(1)}}}{\partial \pi }}
\end{array}
\right) =\left(
\begin{array}{c}
{\frac{\partial \Omega _{_{(1)}}}{\partial q}} \\
{-I}
\end{array}
\right) \;\;\;,\;\;\;R_{_{(1)}}=\left( {\frac{\partial \Omega _{(1)}}
{\partial \alpha }}\right) \;.
\end{equation}
We can see that the matrix $R_{_{(1)}}$, whose elements are $R_{_{(1)}}^{ij}=
{{\displaystyle {\partial \Omega _{_{(1)}}^j \over \partial \alpha _i}}}
={{\displaystyle {\partial ^2\hat{L}(q,\alpha ) \over \partial \alpha _i
\partial \alpha _j}}}$, is the Hessian of $L(q,\dot{q})$ evaluated in
$\dot{q}=\alpha $.

To make apparent when $F_{_{(1)}}$ is regular or singular we can perform the
following manipulation. Introducing the idempotent matrix

\begin{equation}
U_{_{(1)}}=\left(
\begin{array}{ccc}
I & 0 & f^{-1}M_{_{(1)}} \\
0 & I & 0 \\
0 & 0 & -I
\end{array}
\right) \;,
\end{equation}
we construct
\begin{equation}
W_{_{(1)}}=U^{^{T}}F_{_{(1)}}U=\left(
\begin{array}{ccc}
f & 0 & 0 \\
0 & 0 & -R_{_{(1)}} \\
0 & R_{_{(1)}}^{^{T}} & \omega _{_{(1)}}
\end{array}
\right) \;,  \label{W1}
\end{equation}
such that $detF_{_{(1)}}=detW_{_{(1)}}$, where the components of the matrix
\begin{equation}
\omega _{_{(1)}}=M_{_{(1)}}^{^{T}}f^{-1}M_{_{(1)}}  \label{w1}
\end{equation}
are the Poisson brackets with respect to $(q,\pi )$ of the primary
constraints $\Omega _{(1)}$. Hence, because $f$ is regular, the regular or
singular character of $F_{_{(1)}}$ is defined by the matrix

\begin{equation}
\Psi _{_{(1)}}=\left(
\begin{array}{cc}
0 & -R_{_{(1)}} \\
R_{_{(1)}}^{^T} & \omega _{_{(1)}}
\end{array}
\right) \;.
\end{equation}
Thus, $F_{_{(1)}}$ will be regular only if the Hessian $R_{_{(1)}}$ is
regular, in which case the symplectic matrix $F_{_{(1)}}^{-1}$ can be
constructed explicitly:
\begin{equation}
\lbrack F_{_{(1)}}]^{-1}=\left(
\begin{array}{ccc}
f^{-1} & -f^{-1}M_{_{(1)}}R_{_{(1)}}^{-1} & 0 \\
-R_{_{(1)}}^{-1}M_{_{(1)}}^{^T}f^{-1} & R_{_{(1)}}^{-1}\omega
_{_{(1)}}R_{_{(1)}}^{-1} & -R_{_{(1)}} \\
0 & R_{_{(1)}} & 0
\end{array}
\right) \;,
\end{equation}
and we can define the generalized bracket
\begin{equation}
\{E(Q_{_{(1)}}),G(Q_{_{(1)}})\}^{*}={\frac{\partial E(Q_{_{(1)}})}{\partial
Q_{_{(1)}}^{^A}}}\;[F_{_{(1)}}]_{_{AB}}^{-1}\;{\frac{\partial G(Q_{_{(1)}})}
{\partial Q_{_{(1)}}^{^B}}\;.}
\end{equation}
For dynamical functions that only depend on the $(q,\pi )$ variables it
reduces to the Poisson bracket. On the other hand, if the Hessian
$R_{_{(1)}} $ is singular the matrix $F_{_{(1)}}$ is also singular and we
cannot construct such a generalized bracket. To overcome this we must take
into account other constraints, as we shall see in the following section.

Here it is convenient to define the classification of the constraints we
will use. We call primary constraints the ones discussed above, which are
related to the transition to a first order theory. Secondary constraints
directly arise from the zero modes of the Hessian of the original
Lagrangian, and finally possible additional constraints are characterized as
tertiary ones.

\section{The secondary Lagrangian}

A non-regular Hessian $R_{_{(1)}}$ implies a singular $F_{_{(1)}}$, which is
the signature of secondary constraints. These can be made explicit by the
application of the $F_{_{(1)}}$ null eigenvectors on the equations of motion
corresponding to the Lagrangian $L_{_{(1)}}(Q_{_{(1)}})$. Thus, if
$F_{_{(1)}}$ has $m<n$ zero modes $v_{_{(1)}}^r$, $r=1,...,m$, we have $m$
secondary constraints:
\begin{equation}
\Omega _{_{(2)}}^r(q,\pi ,\alpha ,\lambda _{_{(1)}})\equiv v_{_{(1)A}}^r
{\frac{\partial H}{\partial Q_{_{(1)}}^{^A}}}=0\;.  \label{SC}
\end{equation}

The zero modes are given by:
\begin{equation}
v_{_{(1)}}^r=(-v_{_0}^rM_{_{(1)}}^{^T}f^{-1},u^r,-v_{_0}^r)\;,
\end{equation}
where the first component corresponds to the variables $(q,\pi )$, and
$v_{_0}^r$ and $u^r$ are the solutions of :
\begin{equation}
v_{_0}^rR_{_{(1)}}=0\;\;\;,\;\;\;u_{_0}^rR_{_{(1)}}=v_{_0}^r\omega
_{_{(1)}}\;.
\end{equation}
Introducing these vectors into Eq. (\ref{SC}) we get
\begin{equation}
\Omega _{_{(2)}}^r=v_{_{0i}}^r{\frac{\partial H}{\partial q_i}}+v_{_{0i}}^r
{\frac{\partial \Omega _{_{(1)}}^i}{\partial q_j}}{\frac{\partial H}{\partial
\pi _j}}+u_{_i}^r{\frac{\partial H}{\partial \alpha _i}\;,}  \label{SC1}
\end{equation}
and using the expressions (\ref{H}) and (\ref{PC}) for $H(q,\pi ,\alpha )$
and $\Omega _{_{(1)}}^i$ respectively, we finally have:
\begin{equation}
\Omega _{_{(2)}}^r=v_{_{0i}}^r\left[ {\frac{\partial \hat{L}(q,\alpha )}
{\partial q_i}}-{\frac{\partial ^2\hat{L}(q,\alpha )}{\partial \alpha
_i\partial q_j}}\alpha _j\right] +u_{_i}^r\Omega _{_{(1)}}^i\;.
\end{equation}
The last term may be disregarded because it is weakly null, and the
remaining one is precisely a Lagrangian constraint evaluated in $\dot{q}
=\alpha $, which is independent of $\pi _i$, and thus ${\frac{\partial
\Omega _{_{(2)}}^r}{\partial \pi }}=0$. In the present schema the Lagrangian
constraints of the original theory appear as secondary ones, a direct
consequence of the singularity of the Hessian of this theory.

Coming back to Eq. (\ref{SC1}), and using the Poisson bracket defined by
$f^{-1}$ for the canonically conjugated variables $(q,\pi )$, the
constraints $\Omega _{_{(2)}}^r$ may be written as:
\begin{equation}
\Omega _{_{(2)}}^r=v_{_{0i}}^r\{\Omega _{_{(1)}}^i,H\}\;.
\end{equation}
This shows that, although the primary constraints are considered in an
analogous way by our procedure and the Dirac one, there is a relevant
difference with respect to the secondary constraints. Our approach leads
directly to a maximal set of linearly independent secondary constraints, in
contrast with the Dirac method where a rather arbitrary linear combination
of the constraints can be considered as a constraint.

Our following step is to incorporate these new constraints into the
Lagrangian $L_{_{(1)}}(Q_{_{(1)}})$, using again a new set of velocity
Lagrange multipliers, $\dot{\lambda}_{_{(2)}}^i$. This gives place to a new
Lagrangian
\begin{equation}
L_{_{(2)}}(q,\dot{q},\alpha ,\dot{\lambda}_{_{(1)}},\dot{\lambda}
_{_{(2)}})=L_{_{(2)}}(Q_{_{(2)}})=\Delta _{_{(2)}}^{^A}\dot{Q}
_{_{(2)}}^{^A}-W(Q_{_{(1)}})\;,
\end{equation}
where $Q_{_{(2)}}=(q,\pi ,\alpha ,\dot{\lambda}_{_{(1)}},\dot{\lambda}
_{_{(2)}})$. The symplectic matrix corresponding to this last Lagrangian is:
\begin{equation}
F_{_{(2)}}=\left(
\begin{array}{ccc}
f & 0 & M_{_{(2)}} \\
0 & 0 & R_{_{(2)}} \\
-M_{_{(2)}}^{^T} & -R_{_{(2)}}^{^T} & 0
\end{array}
\right) \;,
\end{equation}
with
\begin{equation}
M_{_{(2)}}=\left(
\begin{array}{cc}
{\frac{\partial \Omega _{_{(1)}}}{\partial q}} & {\frac{\partial \Omega
_{_{(2)}}}{\partial q}} \\
{\frac{\partial \Omega _{_{(1)}}}{\partial \pi }} & {\frac{\partial \Omega
_{_{(2)}}}{\partial \pi }}
\end{array}
\right) \;\;\;,\;\;\;R_{_{(2)}}=\left(
\begin{array}{cc}
{\frac{\partial \Omega _{_{(1)}}}{\partial \alpha }} & {\frac{\partial
\Omega _{_{(2)}}}{\partial \alpha }}
\end{array}
\right) \;.  \label{M2}
\end{equation}
Proceeding as in the previous step we construct the matrix $W_{_{(2)}}$,
which has a structure analogous to $W_{_{(1)}}$ in Eq. (\ref{W1}), but now
with:
\begin{equation}
\Psi _{_{(2)}}=\left(
\begin{array}{cc}
0 & -R_{_{(2)}} \\
R_{_{(2)}}^{^T} & \omega _{_{(2)}}
\end{array}
\right) \;.
\end{equation}
The components of $\omega _{_{(2)}}$ are the Poisson brackets of the primary
and secondary constraints, $\Omega _{_{(1)}}$ and $\Omega _{_{(2)}}$, with
respect to $(q,\pi )$. Taking into account Eqs. (\ref{w1}) and (\ref{M2}) it
can be written:
\begin{equation}
\omega _{_{(2)}}=M_{_{(2)}}^{^T}f^{-1}M_{_{(2)}}=\left(
\begin{array}{cc}
\omega _{(1)} & \left[ \frac{\partial \Omega _{(2)}}{\partial q}\right] \\
-\left[ \frac{\partial \Omega _{(2)}}{\partial q}\right] ^T & 0
\end{array}
\right) \;.  \label{w2}
\end{equation}

In contrast with the primary Lagrangian stage, the matrix $R_{_{(2)}}$ is
now rectangular. Our main problem is to study when $F_{(2)}$ is singular or
regular, which is defined by $\Psi _{(2)}$. This is considered in the two
following sections.

\section{A regular $\Psi _{(2)}$ matrix: the restricted phase space and the
generalized brackets}

In this case $F_{(2)}$ is regular and hence $F_{_{(2)}}^{-1}$ exists. As the
matrix $\Psi _{_{(2)}}^{-1}$ exists, we can construct the matrix
\begin{equation}
W_{_{(2)}}^{-1}=U_{_{(2)}}F_{_{(2)}}^{-1}U_{_{(2)}}^{^{T}}=\left(
\begin{array}{cc}
f^{-1} & 0 \\
0 & \Psi _{_{(2)}}^{-1}
\end{array}
\right) \;,
\end{equation}
and easily obtain $F_{_{(2)}}^{-1}$ by considering a block decomposition of
$\Psi _{_{(2)}}^{-1}$. In fact, if we write
\begin{equation}
\Psi _{_{(2)}}^{-1}=\left(
\begin{array}{cc}
(\Psi _{_{(2)}}^{-1})_{_{11}} & (\Psi _{_{(2)}}^{-1})_{_{12}} \\
-(\Psi _{_{(2)}}^{-1})_{_{12}}^{^{T}} & (\Psi _{_{(2)}}^{-1})_{_{22}}
\end{array}
\right)
\end{equation}
where $(\Psi _{_{(2)}}^{-1})_{_{11}}$ and $(\Psi _{_{(2)}}^{-1})_{_{22}}$
are $n\times n$ and $(n+m)\times (n+m)$ antisymmetric matrices respectively,
while $(\Psi _{_{(2)}}^{-1})_{_{12}}$ is a $(n+m)\times n$ rectangular
matrix, the expression for $F_{_{(2)}}^{-1}$ becomes
\begin{equation}
F_{_{(2)}}^{-1}=\left(
\begin{array}{ccc}
\;f^{-1}-f^{-1}M_{_{(2)}}(\Psi
_{_{(2)}}^{-1})_{_{22}}M_{_{(2)}}^{^{T}}f^{-1}\; & -f^{-1}M_{_{(2)}}(\Psi
_{_{(2)}}^{-1})_{_{12}}^{^{T}}\; & -f^{-1}M_{_{(2)}}(\Psi
_{_{(2)}}^{-1})_{_{22}} \\
(\Psi _{_{(2)}}^{-1})_{_{12}}M_{_{(2)}}^{^{T}}f^{-1} & (\Psi
_{_{(2)}}^{-1})_{_{11}} & -(\Psi _{_{(2)}}^{-1})_{_{12}} \\
-(\Psi _{_{(2)}}^{-1})_{_{22}}M_{_{(2)}}^{^{T}}f^{-1} & (\Psi
_{_{(2)}}^{-1})_{_{12}}^{^{T}} & (\Psi _{_{(2)}}^{-1})_{_{22}}
\end{array}
\right)
\end{equation}
Observe that in the block $(1,1)$ of $F_{_{(2)}}^{-1}$ a structure quite
similar to the Dirac bracket can be recognized.

This structure becomes more evident if {{\it $\omega _{(2)}$ }}is regular,
in which case one can perform an invertible transformation in order to write
$\Psi _{_{(2)}}$ in a block diagonal form. This is supplied by the
idempotent matrix
\begin{equation}
S=\left(
\begin{array}{cc}
I & 0 \\
-\omega _{_{(2)}}^{-1}R_{_{(2)}}^{^T} & I
\end{array}
\right) \;,
\end{equation}
such that:
\begin{equation}
\Psi _{_{(2)}}\;\rightarrow \;\hat{\Psi}_{_{(2)}}\equiv S^{^T}\Psi
_{_{(2)}}S=\left(
\begin{array}{cc}
\gamma _{_{(2)}} & 0 \\
0 & \omega _{_{(2)}}
\end{array}
\right) \;,
\end{equation}
where $\gamma _{_{(2)}}=R_{_{(2)}}\omega _{_{(2)}}^{-1}R_{_{(2)}}^{^T}$. In
this case $\gamma _{_{(2)}}$ defines the regular or singular character of
$F_{(2)}$. It is worth remarking that $\det \gamma _{_{(2)}}\neq 0$ is
equivalent to the condition of maximal rank for the matrix of the partial
derivatives of the Lagrangian constraints with respect to the coordinates
and the velocities, which warrants the soundness of the constraints\cite{SM}.
Consequently, in a well defined constrained Lagrangian theory $\gamma
_{_{(2)}}$ must be regular. This allows us to obtain a more explicit
expression for $\Psi _{_{(2)}}^{-1}$ and then construct $F_{_{(2)}}^{-1}$
out of the matrix $R_{_{(2)}}$ and $M_{_{(2)}}$
\begin{equation}
F_{_{(2)}}^{-1}=\left(
\begin{array}{ccc}
\ \;\;f^{-1}-f^{-1}M_{_{(2)}}C_{_{(2)}}M_{_{(2)}}^{^T}f^{-1}\;\; &
\;\;-f^{-1}M_{_{(2)}}\omega _{_{(2)}}^{-1}R_{_{(2)}}^{^T}\gamma
_{_{(2)}}^{-1}\;\; & \;\;-f^{-1}M_{_{(2)}}C_{_{(2)}}\;\; \\
-\gamma _{_{(2)}}^{-1}R_{(2)}\omega _{(2)}^{-1}M_{_{(2)}}^{^T}f^{-1} &
\gamma _{(2)}^{-1} & -\gamma _{(2)}^{-1}R_{_{(2)}}\omega _{_{(2)}}^{-1} \\
C_{_{(2)}}M_{_{(2)}}^{^T}f^{-1} & \omega _{_{(2)}}^{-1}R_{_{(2)}}^{^T}\gamma
_{_{(2)}}^{-1} & C_{_{(2)}}
\end{array}
\right) \;,
\end{equation}
where
\begin{equation}
C_{_{(2)}}=\omega _{_{(2)}}^{-1}-\omega _{_{(2)}}^{-1}R_{_{(2)}}^{^T}\gamma
_{_{(2)}}^{-1}R_{_{(2)}}\omega _{_{(2)}}^{-1}\;,
\end{equation}
which satisfies the relation $C_{_{(2)}}R_{_{(2)}}^{^T}=0$, meaning that
$C_{_{(2)}}$ is constituted by the zero modes of $R_{_{(2)}}^{^T}$.
Therefore $F_{_{(2)}}^{-1}$ has a structure analogous to the one that defines
the symplectic form in the Dirac method, but extended to the whole
configuration space, including the auxiliary variables.

Thus, when {{\it $\omega _{(2)}$ }}is{\it \ }regular we can define a
generalized bracket given by:
\begin{equation}
\{E(Q_{_{(2)}}),G(Q_{_{(2)}})\}^{*}={\frac{\partial E(Q_{_{(2)}})}{\partial
Q_{_{(2)}}^{^A}}}\;[F_{_{(2)}}]_{_{AB}}^{-1}\;{\frac{\partial G(Q_{_{(2)}})}
{\partial Q_{_{(2)}}^{^B}}\;,}
\end{equation}
which in particular for the variables $\alpha _i$ and $\lambda _i$ gives:
\begin{eqnarray}
&&\{\alpha _i,\alpha _j\}^{*}=\gamma _{ij}^{-1}\;, \\
&&\{\lambda _{_{(n)}}^k,\lambda _{_{(n)}}^l\}^{*}=C^{kl}\;.
\end{eqnarray}
This case includes the systems with only second class constraints according
to the Dirac method, and the phase space is a uniquely defined subspace of
the primary one.

It is also possible to have $\Psi _{(2)}$ regular with $\omega _{(2)}$
singular. In this case we can redefine the constraints such that we can
rewrite
\[
\omega _{(2)}^{rxr}=\left(
\begin{array}{ll}
\widetilde{\omega }_{(2)}^{sxs} & 0^{(r-s)xs} \\
0^{sx(r-s)} & 0^{(r-s)x(r-s)}
\end{array}
\right) \;,
\]
where $r$ is the dimension and $s$ the rank of $\omega _{(2)}$. From here on
we can follow the preceeding procedure, but with $\widetilde{\omega }_{(2)}$
instead of $\omega _{(2)}.$ This will lead to expressions similar to the ones
already obtained in the case of $\omega _{(2)}$ regular.

\section{A singular $\Psi _{(2)}$ matrix}

When $\Psi _{_{(2)}}$ is singular, $F_{_{(2)}}$ is also singular, which
makes it impossible to get a generalized bracket structure at this level.
This may happen as a consequence of the existence of some still hidden
constraints, which we call tertiary ones, or of a gauge symmetry of the
secondary Lagrangian. In this case $F_{_{(2)}}$ will have the set of
orthonormal zero modes $\{v_{_{(2)}}^{s}\}$
\begin{equation}
v_{_{(2)}}^{s}F_{_{(2)}}\simeq 0\;,\;\;\;\;\;\;\;\;\;s=1,...,\;m^{\prime
}<m\;,  \label{CM}
\end{equation}
where we use the symbol $\simeq $ to signify that this is a weak equation.
The explicit form of $v_{_{(2)}}^{s}$ is
\begin{equation}
v_{_{(2)}}^{s}=(z^{s}M_{_{(2)}}^{T}f^{-1},y^{s},z^{s})\;,
\end{equation}
where $z^{s}$ and $y^{s}$ satisfy
\begin{equation}
z^{s}R_{_{(2)}}^{T}=0\;\;\;,\;\;\;z^{s}\omega _{_{(2)}}+y^{s}R_{_{(2)}}=0\;.
\label{CM1}
\end{equation}

\subsection{Tertiary constraints}

In an analogous way to secondary constraints, we can also have tertiary ones
given by
\begin{equation}
\Omega _{_{(3)}}^{s}\equiv v_{_{(2)0}}^{s}{\frac{\partial H}{\partial
Q_{_{(2)}}}}\equiv z_{(\nu )}^{s}\{\Omega _{(\nu )},H\}+y^{s}\Omega
_{_{(1)}}\;.  \label{TC}
\end{equation}
We have new constraints $\Omega _{_{(3)}}^{s}$ if $z_{(\nu )}^{s}\{\Omega
_{(\nu )},H\}$ are not linear combinations of the $\Omega _{_{(1)}}$ and
$\Omega _{_{(2)}}$. In this case we must iterate the procedure to include
these constraints in a new Lagrangian
\begin{equation}
L_{_{(3)}}(Q_{_{(3)}})=\Delta _{_{(3)}}^{^{A}}\dot{Q}
_{_{(3)}}^{^{A}}-W(Q_{_{(1)}})\;,
\end{equation}
where $Q_{_{(3)}}=(q,\dot{q},\alpha ,\dot{\lambda}_{_{(1)}},\dot{\lambda}
_{_{(2)}},\dot{\lambda}_{_{(3)}})$ .

\subsection{Gauge Symmetries}

When no new constraints arise from the above equation, the remaining zero
modes give rise to the following symmetry transformations for the equations
of motion (see Appendix C)
\begin{equation}
\delta ^sQ_{_{(2)}}=\overline{v}_{_{(2)}}^s\epsilon \;,  \label{T}
\end{equation}
which in terms of the coordinates of the extended configuration space is:
\begin{eqnarray}
&&\delta ^sQ=\epsilon \;z^sM_{_{(2)}}^Tf^{-1}=\epsilon \;z_{_{(\nu
)r}}^s\{\Omega _{_{(\nu )}}^r,Q\}\;, \\
&&\delta ^s\alpha _i=\epsilon \;y^s\;, \\
&&\delta ^s\lambda _{_{(\nu )}}^r=\epsilon \;z_{_{(\nu )}}^s\;.
\end{eqnarray}
This is a symmetry transformation of the secondary Lagrangian, but not
necessarily a symmetry transformation of the original one. One may wonder
under which conditions the above coordinate transformations can be promoted
to symmetries of the original action. To answer this question we can
consider the variation of the action induced by transformation (\ref{T}
), taking into account that
\begin{equation}
z_{(\nu )}^s\{\Omega _{(\nu )},H\}=U_{(\nu )}^s\Omega _{(\nu )}\;\;.
\label{OH}
\end{equation}
Thus we finally get
\begin{equation}
\delta ^sS\simeq \int_{t_1}^{t_2}dt\;\epsilon \;\left[ U_{_{(\nu )}}^s\Omega
_{_{(\nu )}}-y^s\Omega _{_{(1)}}-z_{_{(\nu )}}^s\left( {\frac{\partial
\Omega _{_{(\nu )}}}{\partial \pi }}\pi -\Omega _{_{(\nu )}}\right) \right]
\;.
\end{equation}
The surface term in the case of internal symmetries will cancel only if the
constraints are first order homogeneous functions of the canonical momenta $
\pi $, as happens in the Dirac method. It is worth remarking that it is
possible to add a term to the time derivative of $\epsilon _s$ in order to
cancel the residual terms, which are all of them proportional to the
constraints
\begin{equation}
\partial _{_t}\epsilon \;\rightarrow \;D_t^s\epsilon =\partial _{_t}\epsilon
-z_{_{(\nu )}}^s\left( U_{_{(\nu )}}^s\Omega _{_{(\nu )}}+y^s\Omega
_{_{(1)}}\right) \epsilon \;.  \label{gt}
\end{equation}
Here we have assumed that the zero modes of $R_{_{(2)}}^T$ are
orthonormalized and there is no sum on $s$. In this way, by introducing this
covariant derivative weakly equivalent to the ordinary one, we obtain a
symmetry holding on the whole of the configuration space, whenever Eq.(\ref
{CM}) is strongly satisfied.

The relations (\ref{CM1}) may be fulfilled in the following two situations.

a) $z^s=0$ and $y^s$ is a zero mode of $R_{_{(2)}}$. Given that $R_{_{(1)}}$
is singular, this implies that the matrix $(\frac{\partial \Omega _{_{(2)}}}
{\partial \alpha })$ has a zero mode. In this case $v_{_{(2)}}^s=(0,y^s,0)$,
and does not lead to new constraints but generates the transformation
symmetry
\begin{eqnarray}
&&\delta Q=0 \\
&&\delta \alpha =\epsilon _sy^s \\
&&\delta \lambda _{_{(\nu )}}=0
\end{eqnarray}
under which the Lagrangian variation is:
\begin{equation}
\delta L_{_{(2)}}=\delta \alpha _i\frac{\partial \Omega }{\partial \alpha _i}
\dot{\lambda}+\delta \alpha _i\frac{\partial H}{\partial \alpha _i}=\epsilon
_sv^s(\Omega _{_{(1)}}+R_{_{(2)}}\dot{\lambda})=\epsilon _sv^s\Omega
_{_{(1)}}\;.
\end{equation}

b) $z^s\neq 0$ and so there are non-trivial zero modes of $R_{_{(2)}}^T$. If
$\omega _{_{(2)}}$ is regular and $\gamma _{_{(2)}}$ is singular, the
relations (\ref{CM1} ) become:
\begin{equation}
z^s=y^sR_{_{(2)}}\omega _{_{(2)}}^{-1}\;\;,\;\;z^s\gamma _{_{(2)}}=0\;,
\end{equation}
so that the zero modes can be written
\begin{equation}
v_{_{(2)}}^s=(v^sR_{_{(2)}}\omega
_{_{(2)}}^{-1}M_{_{(2)}}^Tf^{-1},v^s,-v^sR_{_{(2)}}\omega _{_{(2)}}^{-1})\;,
\end{equation}
and the associated constraints are
\begin{equation}
\Omega _{_{(3)}}^s=v_i^s\Omega _{_{(1)}}{}^i+v^sR_{_{(2)}}\omega
_{_{(2)}}^{-1}\{\Omega ,H\}\;.
\end{equation}
The first term is a linear combination of the primary constraints, and the
second one contains a matrix whose components are the Poisson brackets
$\{\Omega _{_{(1)}},H\}$, a linear combination of the secondary constraints,
and $\{\Omega _{_{(2)}},H\}$. The primary and secondary constraints have
already been considered, and thus the tertiary constraints will only
correspond to a given combination of the Poisson brackets $\{\Omega
_{_{(2)}},H\}$. In general, in this case we must carry on and include these
new constraints in a tertiary Lagrangian, by iterating the preceeding
construction. Tertiary constraints may arise only when the set of zero modes
of $\gamma _{_{(2)}}$ is bigger than the set of zero modes of $R_{_{(2)}}$,
and whenever the time evolution of the secondary constraints, $\{\Omega
_{_{(2)}},H\}$, does not reduce to a linear combination of $\Omega _{_{(1)}}$
and $\Omega _{_{(2)}}$. If no new constraint arises, the transformation
generated by (\ref{T}) leaves the equation of motion invariant on the
constrained subspace.

In the simplest situation where both sets of zero modes are equal, we have
$v^sR_{_{(2)}}=0$ and recover the situation described in the preceeding
subsection.

\section{Some examples}

To clarify the use and possibilities of our method, we consider here two
simple examples which have been widely discussed in the literature. Both
present pathological features from the point of view of the Dirac approach.
The first one is a regular Lagrangian, but with terms of higher degree in
the derivatives. This causes a bifurcation in the canonical structure
according to the usual approach, which gives place to a problem with
multivaluated boundary conditions. Our approach allows us to construct a
symplectic formalism with physical sense, which also gives a consistent
quantum theory. The second example contains a first class constraint,
according to the Dirac classification, which has a weak null bracket with
every dynamical function, and thus does not admit a gauge fixing. In this
case our approach permits a deeper understanding of the origin of the
pathology.

\subsection{1.- Lagrangian of higher order time derivatives}

Let us now consider an example with higher order time derivatives, which has
been studied in Ref. \cite{HT}, defined by the Lagrangian
\begin{equation}
L(\dot{q})=\frac 14\,(\dot{q})^4-\frac 12\,k\,(\dot{q})^2  \label{e1}
\end{equation}

The equation of motion in the configuration space is:
\[
\left( 3\dot{q}^2-k\right) \ddot{q}=0
\]
which implies $\dot{q}=cte.$, i.e. it describes a one-dimensional
free-particle.

When a Legendre transformation is used to construct a canonical formalism,
this Lagrangian leads to velocities which are multivaluated functions of the
canonical momenta, making the classical motion unpredictable since at any
time one can jump from one branch of the Hamiltonian to another. We now
analyze this system in the light of our procedure.

Proceeding as proposed in the first section, we introduce the auxiliary
variables $\alpha $ and $p$ to obtain a first order time derivative
Lagrangian $L_{(0)}$,
\begin{equation}
L_{(0)}(\dot{q},p,\alpha )=p(\dot{q}-\alpha )+\frac 14\,\alpha ^4-\frac 12
\,k\,\alpha ^2\;.  \label{e2}
\end{equation}
Its equations of motion reveal the constraint

\begin{equation}
\Omega \equiv \frac{\partial L_{(0)}}{\partial \alpha }=\alpha
^3-\,k\,\alpha -p\;.  \label{e3}
\end{equation}
By incorporating it to the Lagrangian we get
\begin{equation}
L_{(1)}(\dot{q},p,\alpha ,\lambda )=p\dot{q}+(\alpha ^3-\,k\,\alpha -p)\,
\stackrel{.}{\lambda }-H(p,\alpha )\;,  \label{e4}
\end{equation}
where the Hamiltonian is
\begin{equation}
H(p,\alpha )=-\frac 14\,\alpha ^4+\frac 12\,k\,\alpha ^2+\,p\alpha \;.
\label{e5}
\end{equation}
Arranging the coordinates in a four-dimensional vector $Q_{_{(1)}}^A\equiv
(r=q-\lambda ,p,\alpha ,\lambda )$, one easily obtains the symplectic matrix
$F_{_{(1)}}$
\begin{equation}
F_{_{(1)}}=\left[
\begin{array}{cccc}
0 & -1 & 0 & 0 \\
1 & 0 & 0 & 0 \\
0 & 0 & 0 & 3\alpha ^2-k \\
0 & 0 & k-3\alpha ^2 & 0
\end{array}
\right] \;.  \label{e6}
\end{equation}
Its inverse defines a generalized Poisson bracket
\begin{equation}
\{E(Q_{_{(1)}}),G(Q_{_{(1)}})\}^{*}={\frac{\partial E(Q_{_{(1)}})}{\partial
Q_{_{(1)}}^{^A}}}\;[F_{_{(1)}}]_{_{AB}}^{-1}\;{\frac{\partial G(Q_{_{(1)}})}
{\partial Q_{_{(1)}}^{^B}}\;,}  \label{e7}
\end{equation}
so that the equations of motion can be written as

\begin{equation}
\stackrel{.}{Q}_{_{(1)}}^A=\{Q_{_{(1)}}^A,H\}^{*}\;.  \label{e8}
\end{equation}
They lead to the classical solution $\dot{q}(t)=cte$, consistent with the
original Lagrangian dynamics, but in fact the configuration space is now
two-dimensional, without constraints, and therefore we have two degrees of
freedom and we need four boundary conditions. This establishes a difference
with the usual description for a free particle, given by a Lagrangian
quadratic in the velocity. The equations of motion (\ref{e8}) can be
displayed as:

\begin{eqnarray}
\dot{r} &=&\alpha  \nonumber \\
\dot{\lambda} &=&\alpha ^3-k\alpha -p \\
\dot{\alpha} &=&0  \nonumber \\
\dot{p} &=&0  \nonumber
\end{eqnarray}
which implies that the four phase-space coordinates are constants of motion,
and hence $\stackrel{.}{q}$ $=\dot{r}+\dot{\lambda}.$ is also a constant of
motion. But now it is not sufficient to give, for example, the position and
the velocity at a given time:
\begin{eqnarray}
q_0 &=&r+\lambda \;, \\
\dot{q}_0 &=&\alpha ^3+\left( 1-k\right) \alpha -p\;,  \nonumber
\end{eqnarray}
because there is not enough information to determine a point in the
phase-space.

There is another remark that should be made. The phase space coordinates
$Q_{_{(1)}}^A$ are not canonical, in the sense that the brackets of the
fundamental variables are not equal to one or to zero. If we try to
obtain a canonical structure through a change of coordinates it is necessary
to consider a non-linear transformation, such as $\widetilde{Q}
_{_{(1)}}^A\equiv (r=q-\lambda \,,\,p\,,\tilde{\alpha}=\alpha ^3-k\alpha
,\lambda )$. These new coordinates satisfy canonical brackets, but this type
of transformation also introduces bifurcations in the phase space with a
correlated multivaluated Hamiltonian.

Although the $Q_{_{(1)}}^A$ coordinates are not canonical, the dynamics
described in their terms has a symplectic structure. This is given by Eqs.
(\ref{e6}, \ref{e7}), which define the evolution of the system through the
generalized Hamiltonian (\ref{e5}) and the equations of motion (\ref{e8}).
This symplectic structure can be used to construct a quantum theory by using
the canonical procedure given by $\{A,B\}^{*}\rightarrow
-i[\hat{A},\hat{B}]$, where $\hat{A}$ and $\hat{B}$ are the quantum
operators corresponding to the classical functions $A$ and $B$. This
procedure leads to:
\begin{eqnarray}
\lbrack \hat{r},\hat{p}] &=&i\;,  \label{e9} \\
\lbrack \hat{\alpha},\hat{\lambda}] &=&i\left( k-3\hat{\alpha}^2\right)
^{-1}\;,  \label{e10}
\end{eqnarray}
with all the other fundamental commutators null. A representation for the
fundamental operators that satisfies these commutation relations is:
\begin{eqnarray}
\hat{r} &=&r\;\;\;\;\;\;\;\;\;\;,\;\;\;\hat{\alpha}=\alpha \;, \\
\hat{p} &=&-i\frac \partial {\partial r}\;\;\;\;\;,\;\;\;\;\hat{\lambda}
=-i\left( k-3\alpha ^2\right) ^{-1}\frac \partial {\partial \alpha }\;.
\nonumber
\end{eqnarray}

The time derivatives of the fundamental phase space operators are given by:
\begin{eqnarray}
\dot{r} &=&-i[\hat{r},\hat{H}]=\alpha \;,  \nonumber \\
\dot{\alpha} &=&-i[\hat{\alpha},\hat{H}]=0\;, \\
\dot{p} &=&-i[\hat{p},\hat{H}]=0\;,  \nonumber \\
\dot{\lambda} &=&-i[\hat{\lambda},\hat{H}]=-\left( k-3\alpha ^2\right)
^{-1}\left( \,\hat{p}-\,\alpha ^3+\,k\,\alpha \right) =-\alpha +\frac{\,\hat{
p}+2\alpha ^3}{3\alpha ^2-k}\;,  \nonumber
\end{eqnarray}
which implies that $\hat{p}$, $\hat{\alpha}$, $\frac{d\hat{r}}{dt}$ and
$\frac{d\hat{\lambda}}{dt}$ are constants of motion. For the coordinates of
the configuration space, $r$ and $\lambda $, and their velocities we have
the following commutation relationships:
\begin{eqnarray}
\left[ \hat{r},\frac{d\hat{r}}{dt}\right] &=&\left[ \hat{r},\hat{\alpha}
\right] =0\;, \\
\left[ \hat{\lambda},\frac{d\hat{\lambda}}{dt}\right] &=&-i\left( k-3\alpha
^2\right) ^{-1}\left[ \frac \partial {\partial \alpha },\left( -\alpha +
\frac{\,\hat{p}+2\alpha ^3}{3\alpha ^2-k}\right) \right] =\frac i{k-3\alpha
^2}\left( 1+6\alpha \frac{\hat{p}-\alpha \left( \alpha ^2-k\right) }{\left(
3\alpha ^2-k\right) ^2}\right) \;.
\end{eqnarray}

In particular the velocity of the particle
\begin{equation}
\frac{d\hat{q}}{dt}=\frac{d\hat{r}}{dt}+\frac{d\hat{\lambda}}{dt}=\frac{\,
\hat{p}+2\alpha ^3}{3\alpha ^2-k}\;
\end{equation}
is a constant of motion that can take any real value, as is characteristic
for a free particle. But the commutator of the position and the velocity is
here:
\begin{equation}
\left[ \hat{q},\frac{d\hat{q}}{dt}\right] =\left[ \hat{r}+\hat{\lambda},
\frac{\,\hat{p}+2\alpha ^3}{3\alpha ^2-k}\right] =\frac i{3\alpha ^2-k}
\left[ 1+\frac{6\alpha }{\left( 3\alpha ^2-k\right) ^2}\left( \hat{p}+\alpha
\left( 5\alpha ^2-k\right) \right) \right] \neq i\;.
\end{equation}
This implies that the uncertainty relationship between the position and the
velocity is different from the one resulting when the Lagrangian is
quadratic in the velocity. This result clearly shows that we have a new
quantum theory, but with the same classical limit than the quadratic one.

\subsection{2.- Lagrangian with a pathological Dirac constraint}

The following Lagrangian was proposed by Cawley:
\begin{equation}
L=\dot{q}_1\dot{q}_2+q_3q_1^2\;.
\end{equation}
The construction of a canonical formalism for this system is cumbersome.
This and other related examples have motivated several discussions and ad
hoc modifications of the Dirac method. The Lagrangian equations of motion
give:
\begin{eqnarray}
\stackrel{.}{q}_1 &=&arbitrary\ constant\;,  \nonumber \\
q_2 &=&0\;, \\
q_3 &=&arbitrary\ function\ of\ t\;.  \nonumber
\end{eqnarray}
If we try to apply the Dirac method we find a primary constraint, $p_3=0$,
and a secondary one, $q_1^2=0$. This latter has been called ineffective\cite
{GNH} or a fourth class\cite{LL} constraint, for which it is impossible to
construct a gauge fixing constraint. This prevents us from determining a
physical subspace of the phase space with a symplectic structure. To
overcome this difficulty we could try to use a linearized constraint,
$q_2=0$, instead of the secondary one. From an algebraic point of view both
constraints are equivalent, but this last one leads to an additional
constraint, $p_1=0$. Thus we would now have three first class constraints,
$p_3=0$, $q_2=0$ and $p_1=0$, for which it would be necessary to introduce a
gauge fixing. Each gauge fixing would define a configuration characterized
by:
\begin{eqnarray}
&&q_1=a\ given\ constant\;,  \nonumber \\
&&q_2=0\;, \\
&&q_3=a\ given\ function\ of\ t\;,  \nonumber
\end{eqnarray}
and so it would not reproduce the whole solution space of the Lagrangian
equations. For this reason this Lagrangian is considered a pathological one
from the point of view of the Dirac method.

Now we will apply our method to this Lagrangian. To simplify the discussion
we use the shortcut provided by the algebraic elimination of cyclic
variables. The first order Lagrangian is:
\begin{equation}
L_{(0)}=\sum_{i=1}^{3}\pi _{i}\dot{q}_{i}-H\;,
\end{equation}
with
\begin{equation}
H=\sum_{i=1}^{3}\pi _{i}\alpha _{i}-\alpha _{1}\alpha _{2}-q_{3}q_{1}^{2}\;.
\end{equation}
The equations of motion of the $\alpha $ variables give the constraints:
\begin{eqnarray}
&&\Omega _{(1)}^{1}=\frac{\partial L_{(0)}}{\partial \alpha _{1}}=\alpha
_{2}-\pi _{1}\;,  \nonumber \\
&&\Omega _{(1)}^{2}=\frac{\partial L_{(0)}}{\partial \alpha _{2}}=\alpha
_{1}-\pi _{2}\;, \\
&&\Omega _{(1)}^{3}=\frac{\partial L_{(0)}}{\partial \alpha _{3}}=-\pi
_{3}\;.  \nonumber
\end{eqnarray}
We introduce these constraints into the Lagrangian by means of a set of
velocity Lagrange multipliers $\dot{\lambda}$, and thus we obtain the
primary Lagrangian $L_{(1)}$
\begin{equation}
L_{(1)}=\sum_{i=1}^{3}\pi _{i}\,\dot{q}_{i}-\sum_{i=1}^{3}\Omega _{(1)}^{i}\,
\dot{\lambda}_{i}-H\;.
\end{equation}
The Hessian $R_{(1)}$ has a zero mode leading to the secondary constraint:
\begin{equation}
\Omega _{(2)}=\frac{\partial L_{(0)}}{\partial q_{3}}=-q_{1}^{2}\;.
\end{equation}
Again, we incorporate this constraint to the Lagrangian as before:
\begin{equation}
L_{(2)}=\sum_{i=1}^{3}\pi _{i}\,\dot{q}_{i}-\sum_{i=1}^{3}\Omega _{(1)}^{i}\,
\dot{\lambda}_{i}-q_{1}^{2}\,\dot{\eta}-H\;,
\end{equation}
and now the rectangular matrix $R_{(2)}$ is
\begin{equation}
R_{(2)}=\left[ R_{(1)}\mid 0_{3\times 1}\right] \;.
\end{equation}
It has two zero modes satisfying the Eqs. (\ref{CM1}). One of them is
basically the same as in the previous step, $z^{1}=(0,0,1,0)$ , and gives
rise again to the constraint $\Omega _{(2)}$
\begin{equation}
\Omega _{(3)}^{1}=z_{(\nu )}^{1}\{\Omega _{(\nu )},H\}=\frac{\partial H}{
\partial q_{3}}=-q_{1}^{2}=\Omega _{(2)}\approx 0\;.
\end{equation}
The other zero mode, $z^{2}=(0,0,0,1)$, yields the tertiary constraint
\begin{equation}
\Omega _{(3)}^{2}=z_{(\nu )}^{2}\{\Omega _{(\nu )},H\}=-\frac{\partial
\Omega _{(2)}}{\partial q_{1}}\,\frac{\partial H}{\partial \pi _{1}}
=-2\,q_{1}\alpha _{1}\approx 0\;,
\end{equation}
which is weakly zero by virtue of $\Omega _{(2)}$. Hence no new constraints
actually arise because of these last two zero modes and we may conclude that
these zero modes give rise to a symmetry of the equations of motion.

In fact, the symmetries just found correspond to the last set of equations
of motion, which include all the constraints, and do not necessarily
correspond to symmetries of the original Lagrangian. The question is now
whether it is possible to extend them to the original Lagrangian or not. At
this point the symmetry generated by the second zero mode poses a problem.
Observe that if we try to carry the expression of $\Omega _{(3)}^{2}$ into
the form of Eq. (\ref{TC}), the coefficients $U_{(\nu )}^{s}$, given by Eq.
(\ref{OH}), become ill defined on the constrained subspace, thus preventing
the extension of this symmetry to the whole configuration space, as one can
see through the definition of the covariant derivative given by
Eq. (\ref{gt}).

In summary, the symmetry generated by the first zero mode can be extended to
the whole phase space by constructing an adequate covariant derivative,
which enables us to introduce a gauge fixing without altering the original
Lagrangian equations of motion. But this is not the case for the symmetry
generated by the second zero mode, which is strictly valid only in the
constrained subspace. Any additional condition introduced to make the
presimplectic matrix regular only corresponds in fact to a gauge fixing in
this subspace. Therefore, this additional condition will maintain the
equations of motion in the subspace, but will alter them in the whole space,
and leads to a set of equations of motion different to the one given by the
original Lagrangian.

\section{Final Remarks}

In this paper we develop a general theory of singular Lagrangians and their
constraints. It leads to a systematic and purely Lagrangian approach for
constructing a symplectic structure, thus avoiding the use of a not always
well defined Legendre transformation and the rise of many associated
pathologies. Our approach is based on an enlargement of the original
configuration space (the $q$ space), which makes use of two sets of
auxiliary variables. One set allows us to rewrite the original theory as a
first order one (the $\left( \alpha ,\pi \right) $ variables) and the other
contains the Legendre multipliers (the $\lambda $ variables), which we use
to incorporate all the constraints at the level of Lagrangian equations of
motion . These last variables are introduced as velocities in the enlarged
Lagrangian, such that they acquire the role of conjugate coordinates of the
constraints. In this way we have a well defined first order regular dynamics
in an extended configuration space, which warrants that the $q$ coordinates
satisfy exactly the same equations of motion than in the original
Lagrangian. This procedure allows us to disentangle all the mechanisms
hidden in the Dirac-Bergmann approach, and to identify the source of the
pathologies that appear there.

Our construction gives place to three kinds of constraints. In all the cases
we have a set of primary constraints, which resemble the definition of the
canonical momenta. They characterize the enlargement of the configuration
space necessary to pass from the original theory to a first order one. For a
regular Lagrangian they are the only constraints that appear and our
construction stops here. When the original Lagrangian is singular a new set
of constraints appear, which we call secondary constraints. They are
associated to the zero modes of the Hessian of the original theory. Finally,
there may be additional hidden constraints, not directly related to the zero
modes of the Hessian, which we have characterized as tertiary ones.

These constraints can hold in the whole configuration space or in a
constrained subspace. According to this they can be classified in two
categories. One of them corresponds to what can be called strong
constraints, i.e. the set of independent constraints that act on the whole
configuration space and give place to a reduction of the extended space.
They contain the second class constraints of Dirac.

The other category is constituted by the weak constraints, i.e. constraints
that hold only on a constrained subspace. Strictly speaking, the latter are
the generators of symmetry transformations in such a constrained subspace.
Here there are two possibilities, according to whether they close an algebra
or not. In the first case the symmetry can be promoted to be valid on the
whole configuration space by defining a covariant time derivative for the
parameter of the transformation, weakly equivalent to the ordinary one.
Hence we can give a gauge fixing that does not alter the equations of motion
of the original coordinates $q$, and we can construct a symplectic form. In
the second case, when the constraints do not close an algebra, the symmetry
holds only on the constrained subspace and cannot be extended to the whole
space. In this case we can give a gauge fixing in the constrained subspace,
but it will alter the equations of motion of the original coordinates and it
thus becomes impossible to construct a symplectic form without altering the
original equations of motion.

In summary, when there are strong constraints or weak ones that close an
algebra our approach allows us to construct a consistent symplectic
formalism. Our brackets are defined on the whole extended space, and
correspond to the Dirac brackets in the $(q,\pi )$ subspace when they exist.
Furthermore, we can perform this construction even when
there is no well defined Hamiltonian, as in the example discussed in Section
IV-1, which renders our approach more general than the Dirac method.

When there are weak constraints that do not close an algebra, we cannot
construct a meaningful symplectic formalism consistent with the original
equations of motion. The Lagrangians that lead to this case are pathological
within the Dirac method, and also in other approaches.

In contrast with the Dirac method, in our approach the whole set of
constraints is unambiguously defined and there is a clear connection between
the constraints and the symmetries of the Lagrangian. Some of the main
contributions of our work are that it gives a simple and unified framework
for understanding the origin of the pathologies that appear in the
Dirac-Bergmann formalism, which are explained by the impossibility to
promote a given symmetry from the constrained subspace to the whole
configuration space, and that it brings a more general approach for
constructing a symplectic formalism, even when there is no Hamiltonian at
all in a canonical sense. This last point opens the possibility of
overpassing the limitations of the canonical quantization, and performing an
algebraically consistent quantization on the basis of this non-canonical
symplectic algebra.

\section{ Acknowledgments}

This work was partially supported by the Consejo Nacional de Investigaciones
Cient\'{\i}ficas y T\'{e}cnicas, Argentina.

\appendix

\section{The canonical Hamiltonian}

In the very particular case in which all the cyclic variables $\alpha _i$
can be algebraically eliminated, we are finally faced with a Lagrangian of
the form:
\begin{equation}
L_{_{(0)}}(q,\dot{q},\pi )=\pi _i\dot{q}_i-H(q,\pi )
\end{equation}
with the equation of motion given by:
\begin{equation}
f^{ab}\dot{Q}^b=-\frac{\partial H(Q)}{\partial Q^a}
\end{equation}
where $Q^a=(q_i,\pi _i)$ and
\begin{equation}
f=\left(
\begin{array}{cc}
0 & -I \\
I & 0
\end{array}
\right)  \label{sm}
\end{equation}
a symplectic matrix in the $(q,\pi )$ space. This situation corresponds to a
regular Lagrangian, and the equations of motion can be written
\begin{equation}
\dot{A}(Q)=\{A(Q),H(Q)\}
\end{equation}
where $H(Q)$ is the usual canonical Hamiltonian, in terms of the Poisson
brackets
\begin{equation}
\{E(Q),G(Q)\}={\frac{{\partial E(Q)}}{{\partial Q^a}}}\;f_{ab}^{-1}\;
{\frac{{\partial G(Q)}}{{\partial Q^b}}\;.}
\end{equation}

\section{The Dirac Hamiltonian}

Let us consider a Lagrangian bilinear in the velocities:
\begin{equation}
L=\frac 12m_{ij}\dot{q}_i\dot{q}_j-V(q)\qquad \qquad i=1,...,n\;,
\end{equation}
where the matrix $m_{ij}$ is singular. The procedure presented in Section II
produces the first order Lagrangian:
\begin{equation}
L_{(0)}=\pi _i\dot{q}_i-\pi _i\alpha _i+H\;,
\end{equation}
with
\begin{equation}
H=\frac 12m_{ij}\alpha _i\alpha _j-V(q)\;,
\end{equation}
which leads to the following primary constraints:
\begin{equation}
\Omega _{(1)}^i=m_{ij}\alpha _j-\pi _i\;.
\end{equation}
We assume that the range of $m_{ij}$ is $r<n$, and introduce the operator $U$
that diagonalizes the mass matrix
\begin{equation}
(U^TmU)_j=\lambda _i\delta _{ij}\;,
\end{equation}
and transforms the coordinates according to
\begin{eqnarray}
\alpha \; &\rightarrow &\;\tilde{\alpha}=U^T\alpha U\;, \\
\pi \; &\rightarrow &\;\tilde{\pi}=U^T\pi U\;.  \nonumber
\end{eqnarray}

After this transformation the constraints become
\begin{eqnarray}
\tilde{\pi}_s &=&\lambda _s\tilde{\alpha}_s\qquad \qquad s=1,...,r \\
\tilde{\pi}_p &=&0\qquad \qquad \qquad p=r+1,...,n  \nonumber
\end{eqnarray}
We can use these constraints to eliminate the first $r$ cyclic variables
$\alpha _s$, but the last $n-r$ ones $\alpha _p$ will remain arbitrary. In
such a way the Hamiltonian becomes:
\begin{equation}
\tilde{H}=\frac 12\lambda _s^{-1}\tilde{\pi}_s\tilde{\pi}_s+V(q)+\tilde
{\alpha}_p\tilde{\pi}_p\;,
\end{equation}
which in terms of the original variables $(q_s,\pi _s)$ and the arbitrary
functions $\alpha _p$ can be written
\[
H=\frac{\left( m^{-1}\right) _{st}}2\pi _s\pi _t+V(q)+\alpha _p\pi _p\;,
\]
which is the Dirac Hamiltonian.

\section{Constraints and symmetries}

Given a first order Lagrangian:
\begin{equation}
L(Q,\dot{Q})=f(Q)\dot{Q}-H(Q)\;,
\end{equation}
the form of the corresponding equations of motion is:
\begin{equation}
F(Q)\dot{Q}-(\nabla _{_Q}H)(Q)=0\;.
\end{equation}
If the matrix $F(Q)$ has zero modes $v^r$, $v^rF(Q)=0$, from the equations
of motion we obtain an associated set of constraints
\begin{equation}
\Omega ^r(Q)=v^r(\nabla _{_Q}H)(Q)\;.
\end{equation}
The transformation of  coordinates generated by these zero modes
\begin{equation}
\delta Q=\bar{v}^r\epsilon _r\;,
\end{equation}
where $\epsilon _r$ are time-dependent parameters, produces the Lagrangian
variation
\begin{equation}
\delta L=\frac{\delta L}{\delta Q}\delta Q=\epsilon _rv^r(F(Q)\dot{Q}
-(\nabla _{_Q}H)(Q))=\epsilon _r\Omega ^r\;.  \label{LV}
\end{equation}

The relation (\ref{LV}) shows that the Lagrangian is weakly invariant under
displacements in directions orthogonal to the gradient of the potential. In
a Lagrangian formalism the constraints are part of the equations of motion.
In this sense this symmetry is valid on a subspace determined by the
solutions of the equations of motion, and can be called a ''weak symmetry'',
in contrast with the symmetries independent of the solutions of the
equations of motion which can be called ''strong symmetries ''. When the
constraints close an algebra, this symmetry is related to a strong one,
obtained by a simple redefinition of the derivatives of the parameter
$\epsilon $.


\begin{references}
\bibitem{EE}  This is not the only possible suitable definition of the
auxiliary variables. In this article we restrict ourselves to this
particular election. The implications and possibilities of this freedom in
the construction of a symplectic formalism will be discussed elsewhere.
\end{references}
\end{document}